\documentstyle[preprint,aps,amssymb]{revtex}

\def\beq{\begin{equation}}
\def\eeq{\end{equation}}
\def\beqa{\begin{eqnarray}}
\def\eeqa{\end{eqnarray}}

\def\hf{\textstyle{1\over2}}
\def\half{{1\over2}}

\begin{document}
\draft  
\title{An exactly solvable model of a superconducting\\ to rotational phase
transition}
\author{D.J. Rowe, C. Bahri, and W. Wijesundera}
\address{Department of Physics, University of Toronto,
Toronto, Ontario M5S 1A7, Canada}
\date{\today}
\maketitle

\begin{abstract} We consider a  many-fermion model which exhibits a
transition from a superconducting to a rotational phase with variation of a
parameter in its Hamiltonian.
The model has analytical solutions in its two limits due to the presence of
dynamical symmetries.
However, the symmetries are basically incompatible with one another which
means that no simple solution exists in intermediate situations.
Exact (numerical) solutions are nevertheless possible on a computer
and enable one to study the behaviour of competing but incompatible symmetries
and the phase transitions that result in a semi-realistic situation.
In spite of the complexity of the mixed-symmetry dynamics, the
results are remarkably simple and shed light on the nature of phase transitions.
\end{abstract}



Much can be learned about the validity of approximate many-body theories by
applying them to exactly solvable models \cite{Lip,CBR90}.
We consider a model
with two simple limits (phases): one  superconducting; the other rotational.
Systems of this kind are prevalent in nuclear physics.
Indeed, most open-shell nuclei are regarded as
being vibrational and superconducting or deformed and rotational.
Some nuclei appear to be both rotational and superconducting.
Thus, we consider a model with Hamiltonian
\beq H = \alpha H_1 + (1-\alpha) H_2 \, , \label{eq:1}\eeq
where $H_1$ and $H_2$ are, respectively, Hamiltonians for rotational motion and
superconductivity; $\alpha$ is a parameter in the
range $0\leq \alpha\leq 1$.
With variation of $\alpha$, the system undergoes a
transition from one phase to the other.

A common approach to understanding such systems is with the
 Hartree-Fock-Bogolyubov approximation or one of its phenomenological
variations.    Such mean-field theories yield many solutions in
general: some spherical and
superconducting; others deformed and rotational.
Moreover, with variation of the particle number or parameters in the
Hamiltonian,
phase transitions occur in the solution which lies lowest in energy.
This motivates one to inquire if the phase transitions are real or
an artifact of  the HFB constraints.
When improved  approximations are made, e.g., by restoring
broken rotational and particle number symmetries by angular momentum and number
projection, it appears that the phase transitions are smoothed out \cite{RS80}.
Remnants may survive and even become sharp in large systems.
But, one needs to compare the results with exact calculations,  to know what
they have to say about physical phase transitions.

Exact results for truncated spaces can be obtained with a
shell model code.
For example, Bahri {\it et al.}\ \cite{BED95} have investigated the competition
between superconductivity and rotational dynamics  for a system of identical
nucleons  in the nuclear $fp$ shell.
The remarkable fact is that superconductivity and rotations, separately, can be
treated  rather well as a result of their dynamical symmetries \cite{Rev}.
However, for a combination of the two, one needs a huge space to include all the
relevant correlations.
This is because the dynamical symmetries for rotations and
superconductivity are   incompatible according to the following
definition.

If a Hilbert space ${\Bbb H} $ carries reducible representations of two groups,
the actions of the two groups are said to be {\it incompatible\/} if they do not
commute with one another and if ${\Bbb H} $ has no proper subspace that is
invariant
under both groups.
For example two Lie group actions might be incompatible because there exists no
finite-dimensional Lie group that contains both groups as subgroups.

Microscopic models for the rotational states of an $A$-particle
nucleus are given by Hamiltonians of the form
\beq H_{\rm RM} = {1\over 2m} \sum_n^A p_n^2 + \half m\omega^2 \sum_n^A r_n^2 +
V(Q)
\, ,\eeq
where $V(Q)$ is a low order polynomial in the quadrupole moments of the nucleus
\cite{SymMod}.
The building blocks of such a Hamiltonian ($\sum p_n^2$,  $\sum r_n^2$, and
the components of $Q$) are elements of the (noncompact) symplectic algebra
Sp$ (3,{\Bbb R} )$.
Thus, $H_{\rm RM}$ is diagonalizable on a computer (to within round-off
errors).

Models of superconductivity \cite{Rac} are given by
Hamiltonians of the form
\beq H_{\rm SC} = \sum \varepsilon_j a^\dagger_{jm} a_{jm} -
\sum G_{jj'}  a^\dagger_{jm} a^\dagger_{j\bar m} a_{j'\bar m'} a_{j'm'}
\, ,\eeq
where $\{ a^\dagger_{jm} ,a_{jm}\}$ are creation and annihation operators for
fermions of angular momentum $j$ and $z$-component $m$;
$j\bar m$ label the time-reverse of the state labelled by $jm$.
Such a Hamiltonian has a spectrum generating algebra (SGA) comprising a direct
sum of
 SU$ (2)$ (quasi-spin) algebras \cite{Ker}.
Consequently, it can be diagonalized numerically  (if the
number of single particle states is not too large) \cite{CSR95}.

The actions of the Sp$ (3,{\Bbb R} )$ and quasispin  Lie algebras  are
incompatible.
 As a result, it is not possible to
 diagonalize a linear combination of
 $H_{\rm RM}$ and $H_{\rm SC}$ without an unjustified truncation of the space.
It is then an interesting challenge to find approximate methods for
determining the spectrum of such  combinations.

 To  simplify the problem, we restrict the  symplectic model to a single
spherical harmonic oscillator shell.
The  model then reduces to Elliott's model
\cite{JPE58} for which a suitable Hamiltonian is
\beq H_1 = -\chi {\cal Q}\cdot {\cal Q} \, ,\eeq
where ${\cal Q}$ is the SU$ (3)$  quadrupole tensor.
This Hamiltonian has spectrum given by
\beq E(\lambda\mu J) =-\chi C_2(\lambda\mu)  +
3\chi J(J+1) \, , \eeq
where $(\lambda\mu)$ is an SU$ (3)$ highest weight, $J$ is the angular
momentum, and
\beq C_2(\lambda\mu)= 4(\lambda^2+\mu^2+\lambda\mu + 3\lambda+3\mu)\eeq
is the value of the SU$ (3)$
Casimir invariant.

A simplification of $H_{\rm SC}$ is obtained by putting all single-particle
energies equal and setting $G_{jj'} = G$.
The Hamiltonian $H_{\rm SC}$ then reduces to
\beq H_2 = - G S_+ S_- \, ,\eeq
where
\beqa &S_+ = \sum_{j m>0} (-1)^{j+m} a^\dagger_{jm}a^\dagger_{j,-m} \, ,&
\nonumber\\
&S_- = \sum_{j m>0} (-1)^{j+m} a_{j,-m}a_{jm} \, .&\eeqa
This Hamiltonian has analytical solutions because, as shown by
Kerman {\it et al.}\ \cite{Ker},
$S_\pm$ are the raising and lowering operators of a single SU$ (2)$
algebra with
$[S_+,S_-] = 2S_0$ and $[S_0,S_\pm] = \pm S_\pm$.
It has a spectrum given by
\beq  E(N,s) = -Gs(s+1) +
\textstyle {1\over 4} G (N-2s)(N-2s-2) \, .
\eeq

With these simplifications, the dynamical symmetries of $H_1$ and $H_2$ can be
combined.
 The smallest Lie algebra that contains both
SU$ (2)$  and SU$ (3)$, acting
within the space of a many-particle spherical harmonic oscillator shell, is the
compact symplectic algebra   Sp$ (n)$; an algebra of rank $n$ equal to the
number,
$n= {1\over 2}(\nu+1)(\nu+2)$, of orbital single-particle states for the
shell with $\nu$ harmonic oscillator  quanta.
It is spanned by  operators
\beqa & A_{ij} = \displaystyle\sum_{km} (-1)^{{1\over 2} +m}
a^\dagger_{kim}a^\dagger_{kj,-m} \, ,&\nonumber\\
&B_{ij} = \displaystyle\sum_{km} (-1)^{{1\over 2} +m} a_{kim} a_{kj,-m}
\, , &\label{eq:8}\\
& C_{ij} = \half\displaystyle\sum_{km} (a^\dagger_{kim} a_{kjm} -
a_{kjm}a^\dagger_{kim}) \, ,&\nonumber
\eeqa
where $m$ is summed over spin values $\pm 1/2$, $i$ and $j$ run over the range
$1,\ldots ,\nu$, and
$k$ is summed over $\kappa$ fermion types; e.g.,
  $\kappa =2$ for neutrons and protons.

For $n \geq 3$, Sp$ (n)$ contains Elliott's SU$ (3)$ and Kerman's
SU$ (2)$ quasispin algebra, with
\beq S_+ = \hf\displaystyle\sum_i A_{ii} \, ,\;\;
S_0 = \hf\displaystyle\sum_i C_{ii} \, ,\;\;
S_- = \hf\displaystyle\sum_i B_{ii} \, ,\eeq
as subalgebras.
Thus, sp$(n)$ is a SGA for the Hamiltonian of eqn.\
(\ref{eq:1}).
We consider here the smallest Sp$ (n)$ Lie algebra that contains
both SU$ (3)$ and SU$ (2)$,  i.e.,  Sp$ (3)$.
One recalls that ${\frak so(5)}$ symmetry (isomorphic to ${\frak
sp(2)}$) has been used by Zhang and Demler \cite{Zhang} to explore the
competition
between antiferromagnetism and d-wave superconductivity within the
framework of the Hubbard model.

The relevant irrep of Sp$ (3)$, for present purposes, is the one whose
lowest weight state is the fermion vacuum state
$|0\rangle$.
This state satisfies the equations
\beq B_{ij} |0\rangle =0\, ,\quad C_{ij} |0\rangle = -  \delta_{ij} \kappa
|0\rangle \, ,\eeq
which means that it is a lowest weight state for an  Sp$ (3)$ irrep with
 weight  $( - \kappa, - \kappa, - \kappa)$.

Basis states for an irrep with this lowest weight are generated by repeated
application of the Sp$ (3)$ raising operators  $\{ A_{ij}\}$
to the vacuum state $|0\rangle$.
These operators  transform as a basis for a
${\frak u} (3)$ irrep of highest weight $\{ 200\}$.
Thus, an orthonormal basis $\{ |nKJM\rangle\}$ for the desired  Sp$ (3)$
irrep can be constructed, which reduces the subalgebra chain
\beq {\frak sp} (3) \supset {\frak u} (3) \supset {\frak su} (3)
\supset {\frak so} (3) \supset {\frak so} (2) \, ,\eeq
where $n\equiv \{n_1,n_2,n_3\}$, a triple of integers with $n_1\geq n_2\geq
n_3$, is a ${\frak u} (3)$ highest weight and
$JM$ are the usual ${\frak so} (3) \supset {\frak so} (2)$ angular momentum
quantum numbers; $K$ indexes  multiplicities of $J$ in the SU$ (3)$
irrep with highest weight
$(\lambda,\mu) = (n_1-n_2,n_2-n_3)$.

The matrices of such irreps of the  symplectic Lie algebras are
known \cite{VCS,RWB};
explicit expressions will be given in a more complete report
to follow.

For $\kappa=2$, one obtains Sp(3) representations appropriate for two
kinds of nucleon (neutrons and protons) filling the $1p$ shell.
This space is too small; the most deformed
SU$(3)$ irrep it contains is of highest weight $(\lambda\mu) = (4,0)$
whereas the irreps  relevant for the description
of rotational bands in rare-earth nuclei \cite{JWR91} have  $\lambda\approx
70-100$,
$\mu\approx 0-10$.
Such SU(3) irreps are found in the nuclear shell model for two
kinds of nucleon (i.e., $\kappa =2$) filling the $\nu = 4$ and 5 shells.
However, mixed SU(3) calculations in these shells are impossible.
We therefore  generate the desired SU(3) irreps in our model,
in which the most deformed SU(3) irrep  is of highest weight $(\lambda\mu) =
(2\kappa,0)$, by artificially assigning $\kappa$ values in
the  $20-40$ range.

We stress that  our objective in this paper is not to fit any particular data.
It is rather to construct an exactly solvable model
that can be used to test ideas and many-body theories.
It should also be understood that our model can be derived in many ways.
For example, one can regard its Sp$(3)$ algebra as a subalgebra of a
larger Sp$(n)$ Lie algebra using pseudospin ideas following Ginocchio
\cite{Gin}.
In selecting Sp$(3)$, we have been guided  by
 experience with the noncompact Sp$(3,{\Bbb R} )$ algebra \cite{SymMod}, the
Interacting Boson Model \cite{IBM} and the  Fermion Dynamical Symmetry Model
\cite{FDSM}.
In fact, our model contracts to the Interacting
Boson Model as $\kappa\to \infty$.

Results for the excitation energies of the lowest $J =0$,
2, 4, 6, and 8 states are plotted  in Fig.\ 1  for $N = 2\kappa$, $\kappa
= 5$, 10, and 20, and coupling constants $G= 0.05$ and $\chi =0.005 $
(dimensionless units).
 The $N/\kappa$ = 2 ratio corresponds to selecting the same one-third-filled
shell nucleus in each case.
 The remarkable feature of the results is the sharpness of the transition
from the superconducting to the rotational phase.
For all $\kappa$ values shown, the system is
essentially in a superconducting phase for $\alpha < 0.5$ and a rotational
phase for $\alpha >0.6$.
The figure also shows  the expectation values
$\langle S_+S_-\rangle_\alpha$ and $\langle C_2\rangle_\alpha$ as functions of
$\alpha$ for each $J$, where $C_2$ is the
SU$(3)$ Casimir operator.
The results exhibit  the extraordinary purity of the wave functions
on either side of the phase transition.
For  large $\kappa$ values, $\langle S_+S_-\rangle_\alpha$ remains
 close to its $\alpha =0$ value until $\alpha$ approaches a
critical value $\alpha_0$.
For $\alpha >\alpha_0$,   $\langle S_+S_-\rangle_\alpha$  falls rapidly to a
small $J$-independent value.
Similarly, $\langle C_2\rangle_\alpha$ holds closely to its $\alpha =1$ value
until $\alpha$ approaches $\alpha_0$ from above after which it rapidly falls
to a set of equally spaced values for each $J$.

The purity of the wave functions is also indicated by the closeness of the
results for the excitation energies to what one would get in first order
perturbation theory starting from the appropriate limit.
Thus, if one determines  energy levels to first order in $\alpha$, one obtains
\beqa E_J(\alpha) &=& E_J(0) \nonumber\\
&&+\alpha [ G\langle S_+S_-\rangle_0 -
4\chi \langle C_2\rangle_0 +3\chi J(J+1)]
\, .\eeqa
The linear expressions for the excitation energies, shown in
the figure, indicate that the exactly computed results only diverge
substantially from the first order expressions when $\alpha$ is relatively close
to $\alpha_0$.
On the other hand, if one determines energy levels to first order in
$(1-\alpha)$,  one obtains
\beqa &\lefteqn{E_J(\alpha)= E_J(1)  }&\nonumber\\
&\qquad  &  - (1-\alpha)[ G\langle S_+S_-\rangle_1 -
4\chi \langle C_2\rangle_1 -3\chi J(J+1)]
\, .\eeqa

The way in which the vibrational excitation energies drop precipitously just
prior to the phase transition are reminiscent of
 RPA (random phase approximation) results for quadrupole
vibrational excitation energies.
As shown by Thouless \cite{Thoul60}, within the framework of the RPA, the
collapse of a vibrational energy to zero indicates an instability of the
model ground state against vibrational fluctuations and the expectation of a
phase transition to a deformed state.
 The present model, having a fermion foundation is
clearly amenable to analyses along quasiparticle RPA lines.

It is noteworthy that, for large  $\kappa$ and $N=2\kappa$, there is essentially
no region of $\alpha$ for which the model is both rotational and
superconducting.
It appears that once the system starts to exhibit rotational states, the
superconducting pair correlations are quickly suppressed.
One recalls the similarity, noted by Mottelson and Valatin \cite{MV60}, between
the Coriolis force acting on a particle in a rotating frame of reference and the
interaction of a charged particle with a magnetic field.
One knows that superconductivity tends to be excluded by a magnetic field; this
is the Meissner effect.
Mottelson and Valatin suggested that a similar collapse of
superconducting pair correlations should be expected in  rotational nuclei.
However, their expectation was that the collapse would occur above a certain
critical angular momentum.
Thus, we would not have been surprised to obtain coexisting rotations and
superconductivity at low angular momenta.
Indeed, one of our motivations for considering the present model was to study
superconducting flows in rotational nuclei.
Thus, it is of considerable interest to understand what is special about our
model that precludes this occurrence.
[Subsequent calculations reveal that the strong suppression of
superconductivity occurs in our model only when the number of nucleons $N$
is set equal to $2\kappa$, as in the calculations reported here.  We find
that, for $N>2\kappa$, the pair correlations cause a mixing of SU(3) irreps
in a highly coherent way such that the rotational character of the states is
preserved. Results of these calculations will be reported in the more
detailed analysisof the model to follow.]

Outside of a narrow transition region, both phases of the system show a
remarkable resilience to being perturbed.
This we believe is due, in part, to the fact that the two phases are associated
with incompatible dynamical symmetries.
Put another way, the eigenstates of one phase are spread extremely thinly over
the eigenstates of the other.
However, consideration of other models indicates that this alone is not
sufficient for a sharp phase transition to occur.
For example, to see a sharp phase transition, it appears to be important to have
many particles but only two-body interactions.

\acknowledgments

 The authors thank Joe Repka for helpful suggestions.

\begin{figure}
\vspace{7in}
\includegraphics{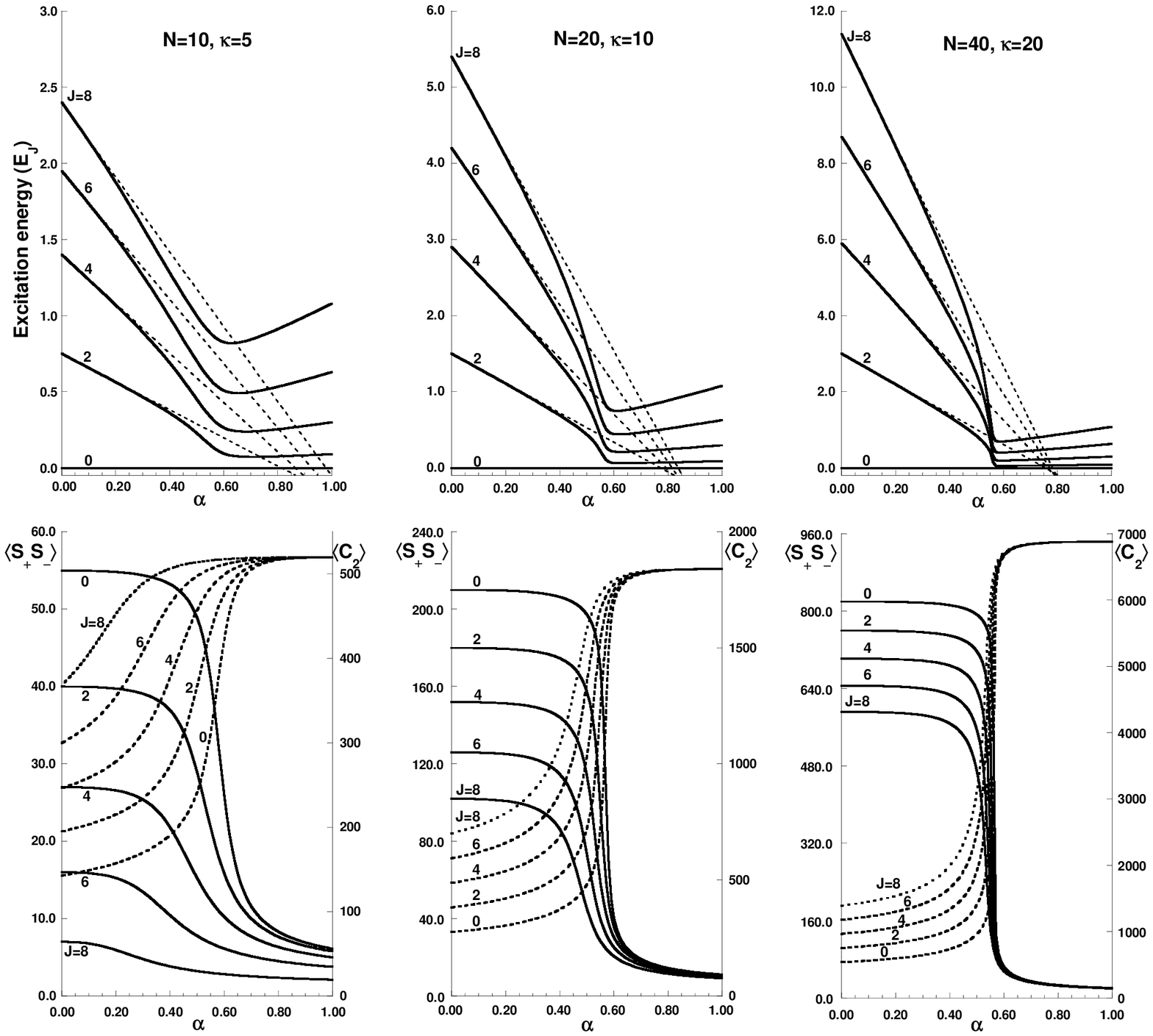}
\caption{The upper figures show the excitation energies for the Hamiltonian of
eqn.\ \protect\ref{eq:1} for different values of $\kappa$ and $N=2\kappa$.
The lower figures show the expectations $\langle S_+S_-\rangle$ and $\langle
C_2\rangle$ as functions of $\alpha$ for the corresponding $\kappa$ values.}
\label{}
\end{figure}

\end{document}